\documentclass[aps,pre,reprint]{revtex4-1}

\usepackage[utf8]{inputenc}
\usepackage{amsmath}
\usepackage{amsfonts}
\usepackage{amssymb}
\usepackage{amsthm}
\usepackage{amsopn}
\usepackage{upgreek}
\usepackage{bm}
\usepackage{epsfig}
\usepackage{nicefrac}

\setcitestyle{round,numbers,sort&compress}
\renewcommand{\eqref}[1]{Eq.~\ref{#1}}
\newcommand{\eqsref}[1]{Eqs.~\ref{#1}}
\newcommand{\figref}[1]{Figure~\ref{#1}}

\newcommand{\refcite}[1]{Ref.~\onlinecite{#1}}

\newcommand{\kB}{k_{\text{B}}}

\begin{document}

\newcommand{\addresscambridge}{Department of Chemistry, University of Cambridge, Lensfield Road, Cambridge CB2 1EW, United Kingdom}

\title{Phase transitions in biological systems with many components}
\author{William M.~Jacobs}
\altaffiliation[Present address:~]{Department of Chemistry and Chemical Biology, Harvard University, 12 Oxford Street, Cambridge, MA, 02138, USA}
\email{wjacobs@fas.harvard.edu}
\affiliation{\addresscambridge}
\author{Daan~Frenkel}
\email{df246@cam.ac.uk}
\affiliation{\addresscambridge}
\date{\today}

\begin{abstract}
Biological mixtures such as the cytosol may consist of thousands of distinct components.
There is now a substantial body of evidence showing that, under physiological conditions, intracellular mixtures can phase separate into spatially distinct regions with differing compositions.
In this paper we present numerical evidence indicating that such spontaneous compartmentalization exploits general features of the phase diagram of a multicomponent biomolecular mixture.
In particular, we show that demixed domains are likely to segregate when the variance in the inter-molecular interaction strengths exceeds a well-defined threshold.
Multiple distinct phases are likely to become stable under very similar conditions, which can then be tuned to achieve multiphase coexistence.
As a result, only minor adjustments to the composition of the cytosol or the strengths of the inter-molecular interactions are needed to regulate the formation of different domains with specific compositions, implying that phase separation is a robust mechanism for creating spatial organization.
We further predict that this functionality is only weakly affected by increasing the number of components in the system.
Our model therefore suggests that, for purely physico-chemical reasons, biological mixtures are naturally poised to undergo a small number of demixing phase transitions.
\end{abstract}

\maketitle

\section*{Introduction}

Biological systems carry out complex chemical reactions involving a large number of interacting components.
However, living cells should not be viewed as well-mixed reaction vessels.
Rather, many of the molecular species in a cell exist as phase-separated domains~\cite{hyman2012beyond,weber2012getting}, and there is considerable evidence that phase separation plays a functional role.
Examples include the formation of cytoplasmic granules~\cite{brangwynne2009germline,elbaum2015disordered,wippich2013dual,molliex2015phase}, nucleoli~\cite{brangwynne2011active,nott2015phase} and amorphous clusters of proteins involved in signaling pathways~\cite{sear2007dishevelled}.
These structures typically appear as liquid-like droplets that are selectively enriched in some components, resulting in spontaneous compartmentalization without the use of membranes~\cite{keating2012aqueous}.
In some cases, more than two compositionally distinct domains, such as the concentric droplets observed in the nucleolus~\cite{feric2016coexisting}, can form simultaneously.
Similarly, in some lipid membranes, interactions between embedded components are believed to drive the formation of phase-separated rafts~\cite{lingwood2010lipid,lingwood2008plasma,kaiser2009order}.
For many of these examples, there is strong evidence that phase separation is essentially an equilibrium phenomenon: active processes, such as ATP hydrolysis, are not necessary to observe spatial segregation~\cite{berry2015rna,weber2015inverse}.

If the formation of compositionally distinct domains is important for the normal functioning of a cell, it is logical to ask what conditions are required to achieve a heterogeneous spatial distribution of components.
To address this question, we first propose three general criteria for `biologically functional' phase separation.
For spontaneous compartmentalization to be useful, each domain should be enriched in a relatively small number of functionally related components.
In order to regulate the formation of domains, stabilizing a phase-separated region should only require small adjustments to the abundances and inter-molecular interactions that contribute to the segregated phase.
Finally, the desired phase behavior should be robust to fluctuations in the interactions between functionally unrelated components.

To understand the physical requirements for meeting these criteria, we consider what happens in a multicomponent mixture with \textit{random} pairwise interactions between $N$ distinct components.
Adopting a model first proposed by Sear and Cuesta~\cite{sear2003instabilities}, we treat the interactions between the molecules in the mixture as pairwise additive and isotropic, i.e., independent of the relative orientations of the molecules~\cite{jacobs2014phase}.
As a result, a single random variable can be used to describe the average strength of the interaction between any pair of components.
Assuming no detailed knowledge of the molecular interactions, we draw each of the ${N (N + 1) / 2}$ random interactions independently from the same distribution (that we shall specify later).

We find that this random-mixture model generates phase diagrams that meet our criteria for biological functionality under rather general conditions.
The phase diagrams of random mixtures fall into two classes, depending on the distribution of interactions and the number of components: one supporting demixed phases that are enriched in a relatively small number of components, and a much simpler case where two phases with nearly identical compositions coexist.
As predicted by a previous simulation study~\cite{jacobs2013predicting}, we demonstrate that increasing the number of components tends to favor the simpler binary phase behavior.
However, we find that demixed phases appear suddenly at a threshold that is primarily determined by the variance of the inter-molecular interactions and scales weakly with the number of components.
Under the conditions where demixing occurs, we find that a limited number of phases, each enriched in a different subset of components, tend to appear at very similar chemical potentials.
This observation implies that, with small changes to the component concentrations, it is possible to turn individual phases on and off and to stabilize multiple phases simultaneously; at the same time, random concentration fluctuations are unlikely to drive demixing into phases with arbitrary compositions.
Our model therefore suggests that biologically functional phase separation is easily achieved in multicomponent mixtures and that the selection of specific phases can be accomplished by locally tuning the inter-molecular interactions.
We provide an intuitive explanation for these results and then discuss the implications for biological mixtures.

\section*{Theory and Monte Carlo Simulations}

\begin{figure*}
  \includegraphics{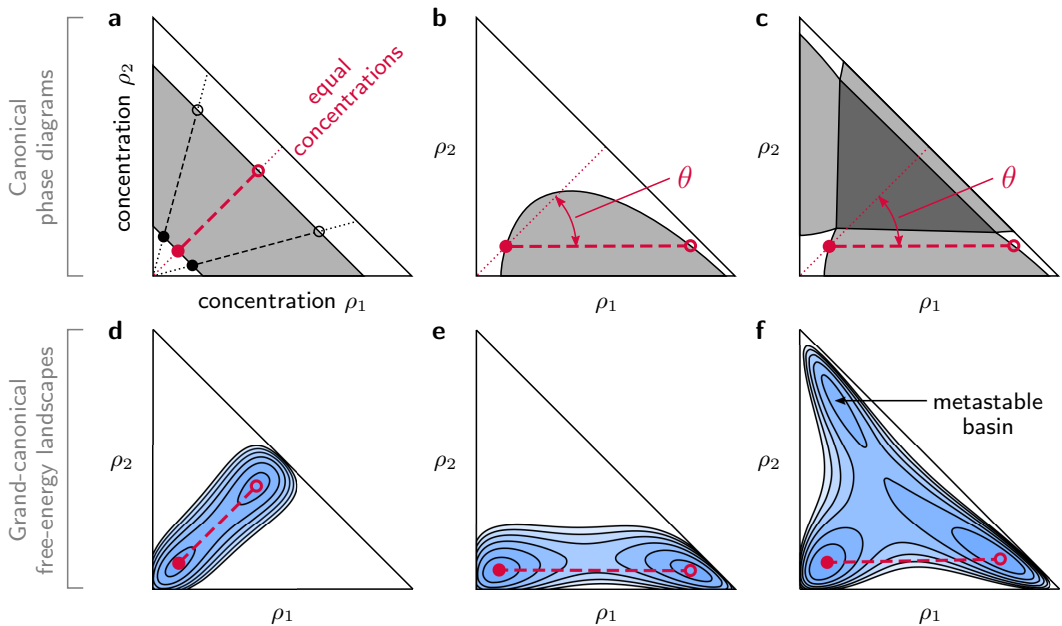}
  \caption{Representative phase diagrams and free-energy landscapes of two-component mixtures.  In the constant-temperature and pressure phase diagrams drawn in panels \textbf{a}--\textbf{c}, single phase regions are shown in white, two-phase coexistence regions in light gray and a three-phase coexistence region in dark gray.  The components have concentrations $\rho_1$ and $\rho_2$, dotted lines indicate constant compositions, and dashed lines indicate example tie lines connecting coexisting phases (circles).  The red dotted line indicates an equimolar parent composition, while the red dashed line indicates the tie line at the boundary of the equimolar homogeneous phase.  The angle of phase separation, $\theta$, is the angle between the highlighted parent composition and the tie line at the cloud point (filled circles).  For each phase diagram, the corresponding panel \textbf{d}--\textbf{f} depicts the free-energy landscape for a mixture with a parent concentration lying on the highlighted tie line.  In these landscapes, the chemical potentials of the components are fixed, and the component concentrations fluctuate among two or more free-energy basins.  In panels \textbf{c} and \textbf{f}, the highlighted tie line is in a two-phase region, but the appearance of a metastable phase on the free-energy landscape indicates close proximity to a three-phase coexistence region.}
  \label{fig:phase_diagrams}
\end{figure*}

\subsection*{Phase diagrams in many dimensions}

Classical thermodynamics predicts that in a mixture with $N$ components, it is possible to observe as many as ${N + 2}$ coexisting equilibrium phases~\cite{gibbs1906scientific}.
The precise number of coexisting phases that appear in any particular system depends on the interactions between the components.
In what follows, we shall use an `implicit solvent' picture in which the solvent is not considered to be a component but is merely the medium in which the various molecular species move.
Moreover, we shall assume that both the temperature and pressure are constant, although, in general, the compositions of the coexisting phases are both temperature and pressure-dependent.

To illustrate some of the features of multicomponent phase separation, we sketch three possible phase diagrams for a two-component mixture, i.e., two solutes plus solvent, in \figref{fig:phase_diagrams}a--c.
Because the components may have dissimilar pairwise interactions, phase separation can drive the formation of phases with differing compositions.
It is convenient to characterize an $N$-component mixture by the concentration vector ${\bm{\rho} \equiv \{\rho_1,\rho_2,\cdots\!\,,\rho_N\}}$, where the total concentration of a phase is ${\phi \equiv \sum_{i=1}^N \rho_i}$.
We use dimensionless units for the component concentrations so that all concentration vectors lie within a unit (${N + 1}$)-dimensional simplex, such that ${0 \le \phi \le 1}$.
The composition of a phase refers to the mole fraction of each component, ${x_i \equiv \rho_i / \phi}$.
Consequently, the composition describes the direction (but not the magnitude) of $\bm{\rho}$.

During a phase transition, both the magnitude and the orientation of $\bm{\rho}$ can change.
Let us imagine that, prior to phase separation, the initially homogeneous mixture is described by the parent concentration vector $\bm{\rho^{(0)}\!}$.
After phase separation, coexistence is established between phases with concentration vectors $\bm{\rho^{(1)}}$ and $\bm{\rho^{(2)}}$ and associated total concentrations $\phi_1 < \phi_2$.
In the phase diagrams shown in \figref{fig:phase_diagrams}a--c, the shaded areas denote coexistence regions, and the circles indicate pairs of daughter phases $\bm{\rho^{(1)}}$ and $\bm{\rho^{(2)}}$; the parent concentration vector lies on the dashed tie line connecting the two daughter phases.
To characterize the nature of a particular phase transition, it is useful to define an `angle of phase separation,' the angle between the parent concentration vector and the tie line connecting the two coexisting phases:
\begin{equation}
  \theta \equiv \cos^{-1} \!\left[\frac{\bm{\rho^{(0)}} \cdot \left(\bm{\rho^{(2)}} - \bm{\rho^{(1)}}\right)}{\left\|\bm{\rho^{(0)}}\right\| \left\|\bm{\rho^{(2)}} - \bm{\rho^{(1)}}\right\|}\right]\!.
\end{equation}
This angle serves as an order parameter that indicates the similarity in the compositions of the parent phase and the coexisting daughter phases.

Multicomponent phase separation can result in phases with equal compositions (\figref{fig:phase_diagrams}a), the formation of a high-concentration phase that is enriched in one component (\figref{fig:phase_diagrams}b), or some combination thereof.
In the case where ${\theta \rightarrow 0}$, the coexisting phases differ only in their total concentrations.
This scenario is referred to as a \textit{condensation} phase transition.
Alternatively, if phase separation at an initially low concentration is driven by the \textit{demixing} of a single component, then ${\theta \rightarrow \theta_{N} \equiv \cos^{-1}(N^{-1/2}})$. 
In principle, in mixtures with more than two components, intermediate phase behavior can occur if $\theta$ takes a value between 0 and $\theta_N$; this scenario implies the selective phase separation of many, but not all, components.
\figref{fig:phase_diagrams}c also shows a three-phase coexistence region, where the mixture phase separates into three coexisting phases with concentrations corresponding to the vertices of the dark shaded triangle.
For parent concentrations lying within this region, multiple demixing transitions occur at the same temperature and pressure.
However, only one phase transition is typically encountered at the boundary of the homogeneous phase, unless the interaction matrix is degenerate.

\subsection*{Characterization of phase diagrams via simulation}

Simulations allow us to study the nature of phase separation in the limit of very many components.
Unlike the canonical phase diagrams shown in \figref{fig:phase_diagrams}a--c, we perform simulations in the grand-canonical ensemble, where the chemical potential of each component is held fixed and the number of particles of each type is allowed to fluctuate.
The grand-canonical ensemble is ideal for studying phase transitions because it allows pure phases to occupy the entire simulation volume.
This means that we can observe coexistence without forming thermodynamically unfavorable interfaces between the pure phases.

Because the component concentrations fluctuate, a grand-canonical simulation samples a free-energy landscape in which low-free-energy basins correspond to the phases in a canonical phase diagram.
For each representative phase diagram shown in \figref{fig:phase_diagrams}a--c, a corresponding free-energy landscape is shown below in \mbox{\figref{fig:phase_diagrams}d--f}.
In each case, the illustration shows the landscape for the component chemical potentials that correspond to a parent concentration vector on the highlighted tie line.
Importantly for what follows, simulations in the grand-canonical ensemble reveal metastable free-energy basins even when the corresponding phase is unstable in the canonical phase diagram.
For example, the metastable basin in the asymmetric free-energy landscape shown in \figref{fig:phase_diagrams}f corresponds to a third phase that must be stable at a nearby set of chemical potentials, indicating that the tie line on the canonical phase diagram in \figref{fig:phase_diagrams}c is close to a three-phase coexistence region.

In order to to determine whether a matrix of inter-molecular interactions supports multiphase coexistence, we follow a simulation strategy that allows us to characterize the canonical phase diagram with a limited number of grand-canonical simulations.
We first assume that all components are present in equal amounts in the parent phase.
Then, starting from the low-concentration homogeneous phase, we locate the lowest-concentration phase boundary that intersects the parent composition vector, i.e., ${\bm{\rho^{(1)}} = \bm{\rho^{(0)}}\!}$.
This concentration vector is commonly called the \textit{cloud point} for a given parent composition and is indicated by the filled circles in \figref{fig:phase_diagrams}\mbox{a--c}.
Lastly, we sample the free-energy landscape and identify metastable free-energy basins at the cloud-point chemical potentials to determine the proximity of this phase boundary to a multiphase coexistence region.
These two sets of simulations are described below.

\subsection*{Monte Carlo simulations of binary phase coexistence}

To locate the cloud point associated with an equimolar parent composition, we carried out grand-canonical Monte Carlo (GCMC) simulations~\cite{frenkel2001understanding} of a three-dimensional lattice model.
In this model, lattice sites may either be vacant, representing the non-interacting solvent, or occupied by one of the $N$ components.
The energy, $U$, of a lattice configuration is calculated by summing the interactions between all pairs of particles on nearest-neighbor (n.n.) lattice sites:
{
\setlength\abovedisplayskip{7pt}
\setlength\belowdisplayskip{7pt}
\begin{equation}
  U = \, - \!\!\!\!\!\! \sum_{(u,v) \in \{\text{n.n.}\}} \!\! \left[ \sum_{i=1}^N \sum_{j=1}^N \delta_{C_u i} \, \epsilon_{ij} \, \delta_{C_v j} \right]\!.
\end{equation}
}
The tuples ${(u,v)}$ run over all nearest-neighbor pairs, ${C_{u}}$ is the component index of the particle at lattice site $u$ (${C_{u}=0}$ if the site is vacant), and $\delta$ is the Kronecker~delta.

We employed multicanonical biasing~\cite{berg1992multicanonical} to facilitate rapid crossing of the free-energy barriers that separate low and high-concentration lattice configurations.
In order to locate the cloud point in these simulations, we tuned the component chemical potentials to achieve equal free energies in the low-concentration phase and the most stable high-concentration phase, ${F^{(1)} = F^{(2)}\!}$, while satisfying the imposed composition constraint on the low-concentration phase, ${\bm{x^{(1)}} = \bm{x^{(0)}}}$.
This strategy has been shown to minimize finite-size effects~\cite{buzzacchi2006simulation}.
We note that this approach does not constrain the simulation to sample only from the two phases of interest if other free-energy basins are present.
For computational efficiency, we used a ${L \times L \times L}$ cubic lattice with periodic boundaries and ${L = 6}$.
Complete details of the simulation method are provided in \refcite{jacobs2013predicting}.

\subsection*{Free-energy landscapes near multiphase coexistence}

For each realization of the interaction matrix, we generated approximately 1000 random lattice configurations, with concentration vectors chosen uniformly from the unit simplex of component concentrations.
We then allowed each initial configuration to evolve via the unbiased GCMC algorithm, with the component chemical potentials fixed at the previously determined cloud point, until the concentration fluctuations stabilized.
These GCMC trajectories tend to travel `downhill' on the free-energy landscape, and, without the multicanonical biasing that was used in the previous set of simulations, the probability of escaping from a free-energy basin is extremely small.
The endpoints of these trajectories therefore lie close to the minima of the free-energy basins, which correspond to stable or metastable thermodynamic phases.
We clustered these endpoints in order to determine the number of distinct free-energy basins, and then we found the mean concentration vector of each of these (meta)stable phases by averaging the endpoint concentration vectors within each basin.

\section*{Results}

\subsection*{Simulations of condensation and demixing in mixtures with random pairwise interactions}

To study an ensemble of mixtures with random interactions, we generated many independent realizations of the pairwise interaction matrices ${\bm{\epsilon} \equiv \{\epsilon_{ij}\}}$.
The interactions between pairs of components were assumed to be independent random variables drawn from a Gaussian distribution with mean $\bar\epsilon$ and variance $\sigma^2$.
(We shall discuss the effects of correlated interactions in the context of the mean-field theory presented below.)
We limited the number of components to 64 for computational tractability and then chose the variance of the Gaussian distributions in order to observe both the demixing and condensation limits of the phase behavior.
The mean interaction strength was fixed at ${\bar\epsilon = 1.07 \epsilon_{\text{c}}}$, where ${-\epsilon_{\text{c}} = -0.87 \kB T}$ is the critical bond energy of the one-component lattice gas, $\kB$ is the Boltzmann constant and $T$ is the absolute temperature.
We shall discuss the relevance of the critical temperature and the rationale for choosing this mean interaction strength in the mean-field section below.

For each realization of ${\bm{\epsilon}}$, we used our simulation strategy both to calculate the component chemical potentials at the lowest-concentration phase boundary and then to determine the number of free-energy basins at this point.
We also computed the average composition of the coexisting high-concentration phase at this set of chemical potentials in order to calculate $\theta$.
As noted above, it is highly unusual to observe multiple coincident transitions precisely at this first phase boundary.
However, when we do observe more than two free-energy basins, multiphase coexistence must occur at very similar component chemical potentials.
In such a case, a stronger driving force --- either higher concentrations or stronger average interactions --- is necessary to push the mixture into the multiphase region.
This scenario is consistent with our expectations for functional multiphase coexistence, because phase transitions that are close in chemical-potential space may be easily manipulated by small changes to the component abundances and inter-molecular interactions.

\begin{figure}
  \includegraphics{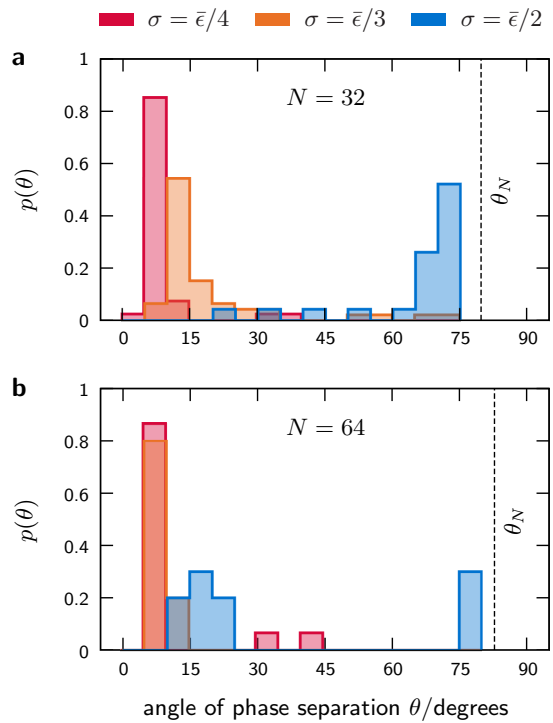}
  \caption{Histograms showing the bimodal distributions of the angle of phase separation in simulations with (a)~32 and (b)~64 components.  The histogram for each random-mixture ensemble was constructed from all high-concentration free-energy basins at the cloud point (see text).  Condensation into two phases with equal compositions is associated with a small angle of phase separation, $\theta$, whereas demixing into phases with dissimilar compositions occurs when $\theta$ approaches $\theta_N$, the angle of phase separation corresponding to the demixing of a single component.  In panel~\textbf{b}, the ${\sigma = \bar\epsilon/2}$ ensemble has nearly equal probabilities of condensation and demixing.}
  \label{fig:distributions}
\end{figure}

\begin{figure*}[t]
  \includegraphics{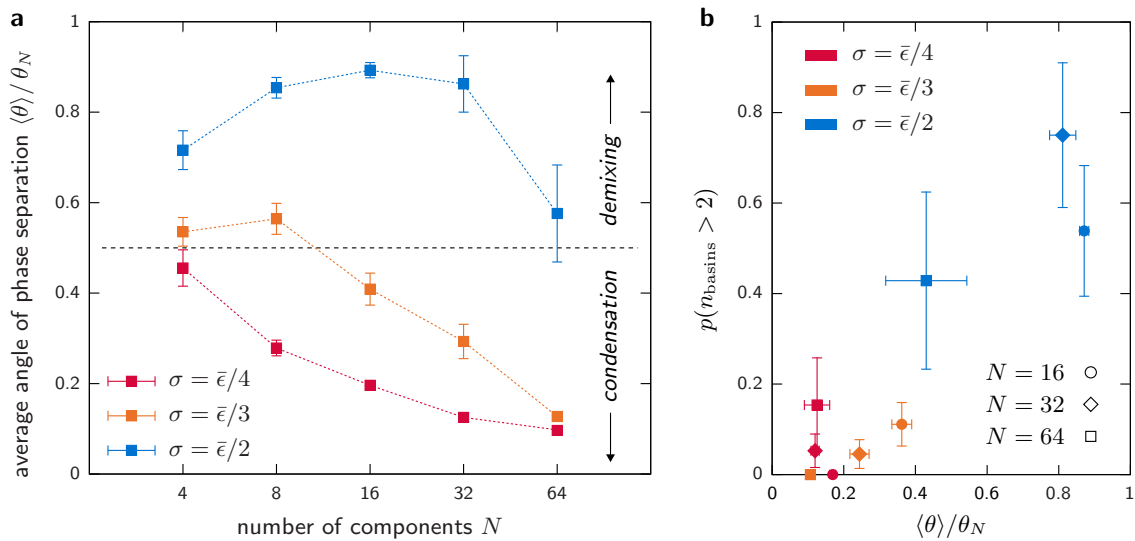}
  \caption{The probability of multiphase coexistence correlates with the angle of phase separation at the cloud point.  (a)~Increasing the number of components in a solution suppresses demixing and leads to a single condensation phase transition, regardless of the variance of the interactions.  Each point represents an average over an ensemble of random mixtures; error bars indicate the standard deviation of the distribution of angles for each ensemble.  The dashed line roughly indicates where the cross-over from demixing to condensation occurs.  (b)~The probability of observing more than two free-energy basins at the cloud point, ${p(n_{\text{basins}} > 2)}$, indicating close proximity to a multiphase coexistence region.  Demixing transitions are associated with the presence of multiple coexisting phases.  Vertical error bars indicate the error due to finite sampling.}
  \label{fig:theta}
\end{figure*}

\subsection*{Bimodal distribution of phase behaviors}

Our simulations show that random interactions result in two distinct types of phase behavior: either a handful of components will demix, or the mixture will condense into two phases with nearly identical compositions.
The resulting bimodal distribution of the angle of phase separation can be seen in \mbox{\figref{fig:distributions}a--b}, where we have included all high-concentration phases that were identified from the free-energy landscapes of dozens of random mixtures with 32 or 64 components.
Low values of $\theta$ correspond to condensation phase transitions.
In this limit, a multicomponent mixture behaves as if all the components were alike.
In contrast, the higher-$\theta$ peak corresponds to demixing phase transitions; in this case, the maximum angle of phase separation is bounded by~$\theta_N$.
Importantly, in simulations with 8 or more components, we rarely find phase behavior that is intermediate between condensation and demixing.

The average phase behavior of a random mixture ensemble depends on both the width of the distribution of inter-molecular interactions, $\sigma$, and the number of components, $N$.
Condensation transitions dominate for large $N$ and small $\sigma$, while demixing transitions are more likely in mixtures with fewer components and a broader distribution of interactions.
In fact, the average angle of phase separation tends toward zero at large $N$, regardless of the variance of the random interactions.
The ensemble-averaged phase behavior is presented in \figref{fig:theta}a, where ${\langle\theta\rangle}$ is shown relative to the largest possible angle of phase separation for a given number of components,~$\theta_N$.
The cross-over between demixing and condensation, where $\langle\theta\rangle$ starts to tend towards 0, occurs at larger~$N$ in random-mixture ensembles with greater values of $\sigma$.
In \figref{fig:theta}a, the cross-over occurs near ${N \simeq 4}$, ${N \simeq 16}$ and ${N \simeq 64}$ for ensembles with ${\sigma = \bar\epsilon/4}$,  ${\sigma = \bar\epsilon/3}$ and ${\sigma = \bar\epsilon/2}$, respectively.
The variance in $\theta$ across realizations of the interaction matrix also decreases as $\langle\theta\rangle$ approaches zero, while the larger variance in the ensemble-averaged angle of phase separation near ${\langle \theta \rangle \simeq \nicefrac{1}{2}}$ is a consequence of the bimodal distribution shown in \figref{fig:distributions}.
We shall discuss the origin of this bimodality and the scaling of $N$, $\sigma$ and $\bar\epsilon$ at the cross-over between demixing and condensation using mean-field calculations below.

\subsection*{Multiphase coexistence correlates with the angle of phase separation}

Free-energy landscapes calculated at the cloud point indicate that multiphase coexistence is most likely to occur in concert with demixing transitions.
In \figref{fig:theta}b, we plot the probability, calculated separately for each random-mixture ensemble, of finding more than one high-concentration phase in the vicinity of the cloud point.
The correlation of the number of free-energy basins identified in our simulations with the ensemble-averaged angle of phase separation implies that the presence of a nearby multiphase coexistence region can be predicted by the angle of phase separation at the cloud point.
This observation allows us to classify the likely phase diagram of a multicomponent mixture based on the statistical distribution of the inter-molecular interactions in a random mixture ensemble.
Furthermore, we find that condensed and demixed free-energy basins only appear together in ensembles where ${\langle \theta \rangle \simeq \nicefrac{1}{2}}$.
This feature indicates that the cross-over from demixing to condensation is a sharp, first-order transition that depends on the relative stability of competing high-concentration phases; the smooth transition in \figref{fig:theta}a is the result of averaging over the bimodal distribution of angles of phase separation.

\subsection*{Mean-field theory of multicomponent\\phase separation}

We can gain further insight into the transition between these two opposing types of phase behaviors with a mean-field analysis of the random-mixture model.
In particular, this approach reproduces the bimodality of the phase behavior observed in our simulations.
This theory also allows us to predict the scaling of the condensation--demixing cross-over as a function of $N$, $\sigma$ and $\bar\epsilon$.

With all components present in equal concentrations in the parent phase, and ignoring spatial correlations between the various particles and vacancies on the lattice, the Helmholtz free energy of a mixture with concentration vector ${\bm{\rho}}$ and lattice coordination number $z$ is
\begin{equation}
  \beta F = \sum_i \rho_i \ln \rho_i + (1 - \phi) \ln (1 - \phi) - \frac{\beta z}{2} \sum_{ij} \rho_i \epsilon_{ij} \rho_j,
  \label{eq:F}
\end{equation}
where ${\phi \equiv \sum_i \rho_i \le 1}$ and ${\beta \equiv (\kB T)^{-1}}$ is the inverse temperature.
In \eqref{eq:F}, the second term accounts for the mixing entropy of vacancies.
Because calculating the total concentration of each phase at coexistence is challenging in the context of a multicomponent mixture, we shall make the simplifying assumption that all high-concentration phases have the same total concentration~$\phi$.
With this approximation, we can write the free-energy difference, ${\Delta F}$, between a phase with an arbitrary composition ${\bm{x}}$ and the equal-composition phase as
\begin{equation}
  \frac{\beta \Delta F}{\phi} = \left( \sum_i x_i \ln x_i + \ln N \right) - \frac{\beta z \phi}{2} \sum_{ij} x_i \Delta\epsilon_{ij} x_j,
  \label{eq:Fx}
\end{equation}
where ${\Delta\epsilon_{ij} \equiv \epsilon_{ij} - \sum_{ij} \epsilon_{ij} / N^2}$.
The first and second terms in \eqref{eq:Fx} represent the differences between the two phases in the mixing entropy and the average interaction energy, respectively.

\begin{figure}[t]
  \includegraphics{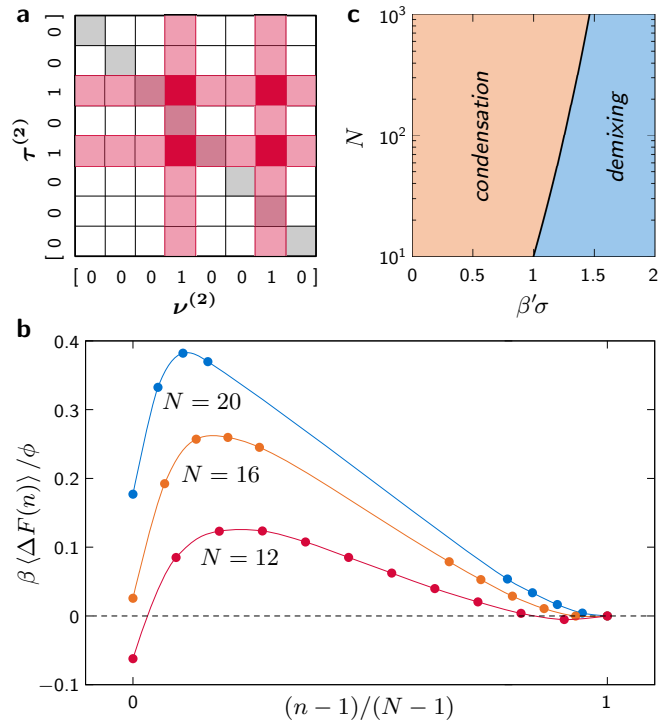}
  \vskip-0.2ex
  \caption{A mean-field theory predicts bimodal phase behavior for random mixtures.  (a)~An illustration of a $2\times2$ sub-matrix (red squares) selected by the vectors ${\bm{\tau^{(2)}}}$ and ${\bm{\upsilon^{(2)}\!}}$.  (b)~The mean-field free-energy difference between the most stable phase comprising ${n^2}$ distinct interaction energies and the equal-composition phase with $N$ components, under the constraint that both phases have the same overall concentration~$\phi$.  Numerical evaluations of \eqref{eq:Fn} are averaged over many realizations of the random matrix ${\bm{\Delta \epsilon}}$; for this illustration, we have chosen ${\beta'\!\sigma = 1.05}$.  The conditions for the cross-over regime can be found by tuning $N$ and $\beta'\!\sigma$ to equalize the free energies of the most stable demixed phase (${n \simeq 1}$) and the condensed phase (${n \simeq N}$).  (c)~The mean-field phase diagram as a function of the control parameters $N$ and ${\beta'\!\sigma}$.  The coexistence curve between condensation and demixing scales approximately as ${\sqrt{\ln N} \sim \beta'\!\sigma}$.}
  \label{fig:mean_field}
\end{figure}

Now let us assume that the components in an arbitrary phase interact via a specific subset of $n^2$ interaction energies and that each of these interactions is equally likely to appear at any bond between nearest-neighbor particles.
Choosing such a subset entails reshuffling the rows and columns of the interaction matrix in order to select a ${n \times n}$ sub-matrix of interaction energies, as illustrated in \figref{fig:mean_field}a.
For example, a phase with ${n = 1}$ could correspond to a single-component phase, if the chosen interaction is on the diagonal of the matrix, or a two-component phase in which the components occupy alternating lattice sites, if the chosen interaction is off-diagonal.
The lowest free energy of a phase with $n^2$ interactions, relative to the equal-composition phase with $N^2$ interactions, is
\begin{equation}
  \!\frac{\Delta F(n)}{\beta^{-1}\phi} = \ln \! \left(\frac{N}{n}\right) - \beta' \!\max\!\left[ \frac{1}{n^2} \! \sum_{ij} \tau^{(n)}_i \! \Delta\epsilon_{ij} \upsilon^{(n)}_j \!\right]_{\!\{\bm{\tau^{(n)}}\!, \bm{\upsilon^{(n)}}\!\}} \!\!\!\!\!\!\!\!\!\!\!\!\!\!\!\!\!\!\!\!\!\!\!\!,\;\;\;
  \label{eq:Fn}
\end{equation}
where ${\beta' \equiv \beta z \phi / 2}$.
The vectors ${\bm{\tau^{(n)}}}$ and ${\bm{\upsilon^{(n)}}\!}$, which have precisely $n$ ones and ${N - n}$ zeros, implement the reshuffling of rows and columns (\figref{fig:mean_field}a).
The second term in \eqref{eq:Fn} is proportional to the maximum value (over all possible vectors ${\bm{\tau^{(n)}}}$ and ${\bm{\upsilon^{(n)}}\!}$) of the \textit{average} of the $n^2$ interaction energies.

The composite parameter $\beta'$ has units of inverse energy and thus sets the scale against which the standard deviation of the interaction distribution should be measured.
This parameter is an implicit function of the mean interaction strength, ${\bar\epsilon}$, which plays an important role in determining the total concentration, $\phi$, at the cloud point.
Although the functional dependence of $\phi$ on the interaction matrix is complicated and best determined through simulation~\cite{jacobs2013predicting}, comparison with a one-component mean-field model indicates that $\beta'$ is closely related to the critical interaction energy, $\epsilon_{\text{c}}$, in a single-component system.
With only one component, it is easy to show that ${\epsilon_{\text{c}} = 4 / \beta z}$ at the critical number density ${\phi_{\text{c}} = \nicefrac{1}{2}}$, meaning that ${\beta'(\epsilon_{\text{c}}) = \epsilon_{\text{c}}^{-1}\!}$.
A condensation phase transition in a multicomponent system cannot occur if ${\bar\epsilon < \epsilon_{\text{c}}}$.
For ${\bar\epsilon > \epsilon_{\text{c}}}$, the single-component analogy indicates that $\beta'$ varies slowly with ${\bar\epsilon - \epsilon_{\text{c}}}$.
We therefore conclude that $\beta'(\bm{\epsilon})$ in multicomponent systems should be interpreted roughly as the critical bond energy for all ${\bar\epsilon \ge \epsilon_{\text{c}}}$.

\subsection*{Mean-field predictions of condensation and demixing}

Numerical analysis shows that \eqref{eq:Fn} reproduces the bimodal phase behavior observed in our simulations.
We evaluated \eqref{eq:Fn} for many realizations of the random interaction matrix and plotted the results in \figref{fig:mean_field}b.
The second term in \eqref{eq:Fn} is minimized when ${n = 1}$, in which case ${\bm{\tau^{(1)}}}$ and ${\bm{\upsilon^{(1)}}}$ pick out the largest entry in the interaction matrix.
As more interaction energies are incorporated, the average of the ${n^2}$ interactions regresses toward the mean, ${\sum_{ij} \Delta \epsilon_{ij} / N^2 = 0}$.
However, this reduction in the average interaction strength is not perfectly balanced by the increase in the mixing entropy.
As a result, \figref{fig:mean_field}b shows that the free-energy difference as a function of $n$ is bistable: either a demixed phase with ${n \simeq 1}$ or a condensed phase with ${n \simeq N}$ is thermodynamically favored.
Equal probabilities for these two scenarios, corresponding to coexistence between demixing and condensation phase transitions, can be obtained by tuning either the total number of components $N$ or the standard deviation of the interaction energies such that ${\Delta F = 0}$ in the ${n \simeq 1}$ phase.

The free-energy curves plotted in \figref{fig:mean_field}b resemble a classical first-order phase transition between stable demixed and condensed phases, where the total number of components and the standard deviation of the interaction strengths are the control parameters.
In random mixtures with uncorrelated interaction energies, high-concentration phases with an intermediate number of components --- corresponding to phases with intermediate values of $\theta$ --- are unlikely to be observed.
The emergence of a condensation phase transition, where ${n \simeq N}$, can be attributed to the greater mixing entropy of the equal-composition, high-total-concentration phase.
In contrast, if ${n \simeq 1}$, the thermodynamic driving force for demixing results from the extreme values of the distribution of random interactions, which implies that only a small number of distinct demixed phases, corresponding to the most strongly interacting component subsets, are likely to be stable.
We find that the fluctuations in the extreme values of $\bm{\Delta\epsilon}$ are large in comparison to the standard deviation of the distribution of interaction energies, leading to a significant variance in ${\Delta F(n)}$ across realizations of the interaction matrix.
As a result, the mean-field theory predicts a smooth cross-over between demixing and condensation behavior when averaged over an ensemble of random mixtures.

We can predict the scaling of the conditions for coexistence between condensation and demixing transitions from the extreme-value statistics of the interaction matrix.
When ${n = 1}$, the expected value of the second term in \eqref{eq:Fn} is ${-\beta'\!\sigma (\sqrt{4.73 \ln N} - 1)}$.
It follows that the coexistence curve between condensation and demixing is ${\sqrt{\ln N} = a \beta'\!\sigma + \sqrt{(a \beta'\!\sigma)^2 - \beta'\!\sigma}}$, where ${a \simeq 1.09}$.
In the large $N$ limit, the resulting scaling relationship ${\sqrt{\ln N} \sim \beta'\!\sigma}$ is considerably steeper than the central-limit-theorem scaling ${\sqrt{N} \sim \sigma/\bar\epsilon}$, which was previously predicted on the basis of a stability analysis of the homogeneous parent phase~\cite{sear2003instabilities}.
Consequently, the $N$--${\beta'\!\sigma}$ phase diagram presented in \figref{fig:mean_field}c suggests that, in mixtures characterized by large values of ${\beta'\!\sigma}$, the number of components required to promote a condensation phase transition may be extremely large.
The simulation results shown in \figref{fig:theta} are also incompatible with ${\sqrt{N} \sim \sigma/\bar\epsilon}$ scaling, although the available data are insufficient to confirm our mean-field prediction.

This mean-field theory can be easily extended to predict the phase behavior of multicomponent mixtures under more general assumptions regarding the inter-molecular interactions and the parent-phase composition.
We have verified that the predicted bistability is robust with respect to weak correlations among the inter-molecular interactions.
In particular, we checked this result for mixtures where the interaction matrix follows the Lorentz--Berthelot mixing rule~\cite{berthelot1898melange} and thus has the form ${\epsilon_{ij} \sim \sqrt{s_i s_j}}$, where the ${\{s_i\}}$ are independent Gaussian random variables with the constraint ${s_i \ge 0}$, and the largest value of ${\epsilon_{ij}}$ is always found on the diagonal of the interaction matrix.
In cases where the interaction distribution is bimodal, our predictions can be applied to the strongly interacting components, which may condense or demix independently from the other weakly interacting components.
In mixtures with unequal component concentrations, \eqsref{eq:F} and \ref{eq:Fx} must be amended to account for the decrease in entropy relative to the parent phase, ${\sum_i x_i \ln x_i^{(0)}\!}$.
The free-energy difference for each demixed phase will then depend on the parent composition as well as the selected subset of inter-molecular interactions.

\section*{Discussion}

For phase separation to be a viable mechanism of spatial organization, it must be possible to tune the stability of multiple coexisting phases via a limited number of local parameters.
In addition, the phase behavior of a mixture with many components should be insensitive to fluctuations in the interactions between each and every pair of components.
We have shown that a model in which the inter-molecular interactions are chosen randomly produces phase diagrams that satisfy this requirement under rather general conditions.
The most probable phase behavior of such a mixture falls into one of two classes, depending on the number of components and the statistical distribution of the inter-molecular interactions.
Under the conditions where a homogeneous mixture first becomes unstable, either a few components will demix, or all components will condense together.
The former type of phase behavior typically supports coexistence among multiple compositionally distinct phases, given small adjustments to the component chemical potentials.
This means that, when demixing occurs in a random mixture, the conditions for multiphase coexistence are both readily accessible and easily tunable.

Our results have a number of important implications for biological mixtures.
One consequence of bimodal phase behavior is that only a small number of interactions need to be tuned to achieve coexistence among multiple demixed phases.
In the absence of strong correlations among the inter-molecular interactions, our mean-field theory predicts that each demixed phase is unlikely to be enriched in more than a few components.
The stabilities of these phases can therefore be tuned by optimizing a small number of component chemical potentials.
To manipulate the phase behavior within the lifetime of a single cell, such control is likely to be achieved through the regulation of macromolecular abundances, while on evolutionary timescales, the phase behavior may be tuned via local changes to the inter-molecular interactions.
In an intracellular mixture, multiphase demixing results in the formation of compositionally inhomogeneous droplets, such as spontaneous nucleolar compartmentalization~\cite{feric2016coexisting}.
However, the spatial ordering of the phases within a droplet depends on the interfacial free energies, which we have not examined here.

Functional multiphase coexistence furthermore requires that demixing transitions are not hidden by a dominating condensation instability.
Our mean-field theory predicts that the required number of components for condensation grows rapidly with the variance of the interaction-energy distribution.
This scaling implies that demixing transitions can be observed in mixtures that contain a few thousand distinct components with an interaction-distribution standard deviation on the order of a few ${\kB T}$.
In our simulations, we have chosen the variance of the interaction strengths in order to explore the behavior of the model over a computationally tractable number of components.
However, these parameters could instead be obtained from high-throughput protein--protein interaction assays.
A previous analysis of Yeast 2-Hybrid experiments~\cite{zhang2008constraints} found that the proteome-wide distribution of nonspecific interactions is indeed quite broad, with an estimated mean and standard deviation of $-4\kB T$ and $2.5\kB T$, respectively.

Our model also shows that multiphase coexistence is not simply a consequence of having a large number of components.
In contrast to the complexity that might be expected from the Gibbs phase rule, phase separation into two phases with similar compositions is a common outcome for mixtures with heterogeneous interactions.
This observation helps to explain the condensation-like phase behavior that has been observed for components embedded in lipid membranes~\cite{hyman2012beyond}.
Although such simple phase behavior might look like that of a single-component solution, this is not a result of the interacting components being indistinguishable.
On the contrary, this behavior is a consequence of the increased mixing entropy that stabilizes homogeneous-composition phases in mixtures with a large number of distinguishable components.

Owing to the simplicity of our model and the minimal constraints imposed on the form of the interaction matrix, our results can be applied to a wide variety of biological systems, including both two-dimensional membranes and three-dimensional fluids.
However, additional information about the inter-molecular interactions is required before quantitative conclusions can be drawn for any specific system~\cite{deeds2007robust}.
Whereas we have assumed uncorrelated pairwise interactions in our simulations, some classes of biomolecules, such as intrinsically disordered proteins, are more likely to interact promiscuously.
In particular, experiments have implicated proteins with low-complexity sequences in some examples of \textit{in vivo} phase separation~\cite{brangwynne2015polymer}.
These considerations imply a more structured interaction matrix than the uncorrelated Gaussian ensemble.
In addition, protein expression levels are typically anticorrelated with the propensity of a protein to form nonspecific interactions~\cite{zhang2008constraints,heo2011topology,johnson2011nonspecific}.
While our mean-field calculations indicate that bimodal phase behavior is robust to physically realistic correlations among the random inter-molecular interactions, very strong correlations may affect this result.
Further investigation is warranted to examine the effects of correlated interactions and unequal concentrations on the phase behavior of multicomponent mixtures.

In summary, we have described the phase behavior of a minimal model of a biological mixture with many components.
We have shown that the general features of the phase diagrams of such mixtures depend on a small number of parameters that describe the distribution of pairwise interactions between components.
With a sufficiently broad distribution of weakly correlated inter-molecular interactions, these mixtures allow for tunable coexistence among multiple phases.
Our results therefore suggest that multicomponent mixtures with random interactions exhibit the essential thermodynamic features that are required for phase separation to play a functional role in a biological system.
While the study of high-dimensional phase diagrams has long been a topic of theoretical interest~\cite{griffiths1970critical,griffiths1975phase}, recent explorations of the organization of complex biological systems --- which are truly multicomponent mixtures --- have brought new relevance to this problem.
We hope that the general principles developed here will guide future studies in this emerging field.

\begin{acknowledgments}
W.M.J.~acknowledges support from the Gates Cambridge Trust and the National Science Foundation Graduate Research Fellowship under Grant No.~DGE-1143678.
\end{acknowledgments}


\begin{thebibliography}{31}
\providecommand{\url}[1]{\texttt{#1}}
\providecommand{\urlprefix}{ }

\bibitem[Hyman and Simons(2012)]{hyman2012beyond}
Hyman, A.~A., and K.~Simons, 2012.
\newblock Beyond oil and water -- {P}hase transitions in cells.
\newblock \emph{Science} 337:1047--1049.

\bibitem[Weber and Brangwynne(2012)]{weber2012getting}
Weber, S.~C., and C.~P. Brangwynne, 2012.
\newblock Getting {RNA} and protein in phase.
\newblock \emph{Cell} 149:1188--1191.

\bibitem[Brangwynne et~al.(2009)Brangwynne, Eckmann, Courson, Rybarska, Hoege,
  Gharakhani, J{\"u}licher, and Hyman]{brangwynne2009germline}
Brangwynne, C.~P., C.~R. Eckmann, D.~S. Courson, A.~Rybarska, C.~Hoege,
  J.~Gharakhani, F.~J{\"u}licher, and A.~A. Hyman, 2009.
\newblock Germline P granules are liquid droplets that localize by controlled
  dissolution/condensation.
\newblock \emph{Science} 324:1729--1732.

\bibitem[Elbaum-Garfinkle et~al.(2015)Elbaum-Garfinkle, Kim, Szczepaniak, Chen,
  Eckmann, Myong, and Brangwynne]{elbaum2015disordered}
Elbaum-Garfinkle, S., Y.~Kim, K.~Szczepaniak, C.~C.-H. Chen, C.~R. Eckmann,
  S.~Myong, and C.~P. Brangwynne, 2015.
\newblock The disordered {P} granule protein {LAF-1} drives phase separation
  into droplets with tunable viscosity and dynamics.
\newblock \emph{Proc. Natl. Acad. Sci. U.S.A.} 201504822.

\bibitem[Wippich et~al.(2013)Wippich, Bodenmiller, Trajkovska, Wanka,
  Aebersold, and Pelkmans]{wippich2013dual}
Wippich, F., B.~Bodenmiller, M.~G. Trajkovska, S.~Wanka, R.~Aebersold, and
  L.~Pelkmans, 2013.
\newblock Dual specificity kinase {DYRK3} couples stress granule
  condensation/dissolution to {mTORC1} signaling.
\newblock \emph{Cell} 152:791--805.

\bibitem[Molliex et~al.(2015)Molliex, Temirov, Lee, Coughlin, Kanagaraj, Kim,
  Mittag, and Taylor]{molliex2015phase}
Molliex, A., J.~Temirov, J.~Lee, M.~Coughlin, A.~P. Kanagaraj, H.~J. Kim,
  T.~Mittag, and J.~P. Taylor, 2015.
\newblock Phase separation by low complexity domains promotes stress granule
  assembly and drives pathological fibrillization.
\newblock \emph{Cell} 163:123--133.

\bibitem[Brangwynne et~al.(2011)Brangwynne, Mitchison, and
  Hyman]{brangwynne2011active}
Brangwynne, C.~P., T.~J. Mitchison, and A.~A. Hyman, 2011.
\newblock Active liquid-like behavior of nucleoli determines their size and
  shape in {X}enopus laevis oocytes.
\newblock \emph{Proc. Natl. Acad. Sci. U.S.A.} 108:4334--4339.

\bibitem[Nott et~al.(2015)Nott, Petsalaki, Farber, Jervis, Fussner,
  Plochowietz, Craggs, Bazett-Jones, Pawson, Forman-Kay, and
  Baldwin]{nott2015phase}
Nott, T.~J., E.~Petsalaki, P.~Farber, D.~Jervis, E.~Fussner, A.~Plochowietz,
  T.~D. Craggs, D.~P. Bazett-Jones, T.~Pawson, J.~D. Forman-Kay, and A.~J.
  Baldwin, 2015.
\newblock Phase transition of a disordered nuage protein generates
  environmentally responsive membraneless organelles.
\newblock \emph{Mol. Cell} 57:936--947.

\bibitem[Sear(2007)]{sear2007dishevelled}
Sear, R.~P., 2007.
\newblock Dishevelled: {A} protein that functions in living cells by phase
  separating.
\newblock \emph{Soft Matter} 3:680--684.

\bibitem[Keating(2012)]{keating2012aqueous}
Keating, C.~D., 2012.
\newblock Aqueous phase separation as a possible route to compartmentalization
  of biological molecules.
\newblock \emph{Acc. Chem. Res.} 45:2114--2124.

\bibitem[Feric et~al.(2016)Feric, Vaidya, Harmon, Mitrea, Zhu, Richardson,
  Kriwacki, Pappu, and Brangwynne]{feric2016coexisting}
Feric, M., N.~Vaidya, T.~S. Harmon, D.~M. Mitrea, L.~Zhu, T.~M. Richardson,
  R.~W. Kriwacki, R.~V. Pappu, and C.~P. Brangwynne, 2016.
\newblock Coexisting Liquid Phases Underlie Nucleolar Subcompartments.
\newblock \emph{Cell} 165:1686--1697.

\bibitem[Lingwood and Simons(2010)]{lingwood2010lipid}
Lingwood, D., and K.~Simons, 2010.
\newblock Lipid rafts as a membrane-organizing principle.
\newblock \emph{Science} 327:46--50.

\bibitem[Lingwood et~al.(2008)Lingwood, Ries, Schwille, and
  Simons]{lingwood2008plasma}
Lingwood, D., J.~Ries, P.~Schwille, and K.~Simons, 2008.
\newblock Plasma membranes are poised for activation of raft phase coalescence
  at physiological temperature.
\newblock \emph{Proc. Natl. Acad. Sci. U.S.A.} 105:10005--10010.

\bibitem[Kaiser et~al.(2009)Kaiser, Lingwood, Levental, Sampaio, Kalvodova,
  Rajendran, and Simons]{kaiser2009order}
Kaiser, H.-J., D.~Lingwood, I.~Levental, J.~L. Sampaio, L.~Kalvodova,
  L.~Rajendran, and K.~Simons, 2009.
\newblock Order of lipid phases in model and plasma membranes.
\newblock \emph{Proc. Natl. Acad. Sci. U.S.A.} 106:16645--16650.

\bibitem[Berry et~al.(2015)Berry, Weber, Vaidya, Haataja, and
  Brangwynne]{berry2015rna}
Berry, J., S.~C. Weber, N.~Vaidya, M.~Haataja, and C.~P. Brangwynne, 2015.
\newblock {RNA} transcription modulates phase transition-driven nuclear body
  assembly.
\newblock \emph{Proc. Natl. Acad. Sci. U.S.A.} 112:E5237--E5245.

\bibitem[Weber and Brangwynne(2015)]{weber2015inverse}
Weber, S.~C., and C.~P. Brangwynne, 2015.
\newblock Inverse size scaling of the nucleolus by a concentration-dependent
  phase transition.
\newblock \emph{Curr. Biol.} 25:641--646.

\bibitem[Sear and Cuesta(2003)]{sear2003instabilities}
Sear, R.~P., and J.~Cuesta, 2003.
\newblock Instabilities in complex mixtures with a large number of components.
\newblock \emph{Phys. Rev. Lett.} 91:245701.

\bibitem[Jacobs et~al.(2014)Jacobs, Oxtoby, and Frenkel]{jacobs2014phase}
Jacobs, W.~M., D.~W. Oxtoby, and D.~Frenkel, 2014.
\newblock Phase separation in solutions with specific and nonspecific
  interactions.
\newblock \emph{J. Chem. Phys.} 140:204109.

\bibitem[Jacobs and Frenkel(2013)]{jacobs2013predicting}
Jacobs, W.~M., and D.~Frenkel, 2013.
\newblock Predicting phase behavior in multicomponent mixtures.
\newblock \emph{J. Chem. Phys.} 139:024108.

\bibitem[Gibbs(1906)]{gibbs1906scientific}
Gibbs, J.~W., 1906.
\newblock The scientific papers of {J}. {W}illard {G}ibbs, volume~1.
\newblock Longmans, Green and Company, New York.

\bibitem[Frenkel and Smit(2001)]{frenkel2001understanding}
Frenkel, D., and B.~Smit, 2001.
\newblock Understanding molecular simulation: {F}rom algorithms to
  applications.
\newblock Academic Press.

\bibitem[Berg and Neuhaus(1992)]{berg1992multicanonical}
Berg, B.~A., and T.~Neuhaus, 1992.
\newblock Multicanonical ensemble: {A} new approach to simulate first-order
  phase transitions.
\newblock \emph{Phys. Rev. Lett.} 68:9.

\bibitem[Buzzacchi et~al.(2006)Buzzacchi, Sollich, Wilding, and
  M{\"u}ller]{buzzacchi2006simulation}
Buzzacchi, M., P.~Sollich, N.~B. Wilding, and M.~M{\"u}ller, 2006.
\newblock Simulation estimates of cloud points of polydisperse fluids.
\newblock \emph{Phys. Rev. E} 73:046110.

\bibitem[Berthelot(1898)]{berthelot1898melange}
Berthelot, D., 1898.
\newblock Sur le m{\'e}lange des gaz.
\newblock \emph{Compt. Rendus} 126:1703--1706.

\bibitem[Zhang et~al.(2008)Zhang, Maslov, and
  Shakhnovich]{zhang2008constraints}
Zhang, J., S.~Maslov, and E.~I. Shakhnovich, 2008.
\newblock Constraints imposed by non-functional protein--protein interactions
  on gene expression and proteome size.
\newblock \emph{Mol. Sys. Biol.} 4:210.

\bibitem[Deeds et~al.(2007)Deeds, Ashenberg, Gerardin, and
  Shakhnovich]{deeds2007robust}
Deeds, E.~J., O.~Ashenberg, J.~Gerardin, and E.~I. Shakhnovich, 2007.
\newblock Robust protein--protein interactions in crowded cellular
  environments.
\newblock \emph{Proc. Natl. Acad. Sci. U.S.A.} 104:14952--14957.

\bibitem[Brangwynne et~al.(2015)Brangwynne, Tompa, and
  Pappu]{brangwynne2015polymer}
Brangwynne, C.~P., P.~Tompa, and R.~V. Pappu, 2015.
\newblock Polymer physics of intracellular phase transitions.
\newblock \emph{Nat. Phys.} 11:899--904.

\bibitem[Heo et~al.(2011)Heo, Maslov, and Shakhnovich]{heo2011topology}
Heo, M., S.~Maslov, and E.~I. Shakhnovich, 2011.
\newblock Topology of protein interaction network shapes protein abundances and
  strengths of their functional and nonspecific interactions.
\newblock \emph{Proc. Natl. Acad. Sci. U.S.A.} 108:4258--4263.

\bibitem[Johnson and Hummer(2011)]{johnson2011nonspecific}
Johnson, M.~E., and G.~Hummer, 2011.
\newblock Nonspecific binding limits the number of proteins in a cell and
  shapes their interaction networks.
\newblock \emph{Proc. Natl. Acad. Sci. U.S.A.} 108:603--608.

\bibitem[Griffiths and Wheeler(1970)]{griffiths1970critical}
Griffiths, R.~B., and J.~C. Wheeler, 1970.
\newblock Critical points in multicomponent systems.
\newblock \emph{Phys. Rev. A} 2:1047.

\bibitem[Griffiths(1975)]{griffiths1975phase}
Griffiths, R.~B., 1975.
\newblock Phase diagrams and higher-order critical points.
\newblock \emph{Phys. Rev. B} 12:345.

\end{thebibliography}
\end{document}